\documentclass[showkeys]{revtex4}
\usepackage{graphicx}
\usepackage{times}
\usepackage{amsmath}  

\begin{document}

\title{Maximum mass of magnetic white dwarfs}

\author{D. Manreza Paret}
\email{dmanreza@fisica.uh.cu}
\affiliation{Facultad de F{\'i}sica, Universidad de la Habana, San L{\'a}zaro y L, Vedado La Habana, 10400, Cuba}
\author{A. Perez Martinez}
\email{aurora@icimaf.cu}
\affiliation{Instituto de Cibern{\'e}tica Matem{\'a}tica y F{\'i}sica (ICIMAF) Calle E esq 15 No. 309 Vedado\\
La Habana, 10400, Cuba}
\author{J.~E.~Horvath}
\email{foton@iag.usp.br}
\affiliation{Instituto de Astronomia, Geofisica e Ciencias Atmosfericas USP, Rua do Matao 1226, 05508-900 Sao Paulo SP, Brazil}

\date{\today}

\begin{abstract}
We revisit in this work the problem of the maximum masses of magnetized White
Dwarfs (WD). The impact of a strong magnetic field onto the structure equations is addressed. The
pressures become anisotropic due to the presence of the magnetic field and split into a
parallel and perpendicular components. We first construct stable solutions of TOV equations for the parallel pressures, and found that physical solutions vanish for the perpendicular pressure when $B \gtrsim 10^{13}$~G. This fact establishes an upper bound for a magnetic field and the stability of the configurations in the (quasi) spherical approximation. Our findings also indicate that it is not possible to obtain stable magnetized WD with super Chandrasekhar masses because the values of the magnetic field needed for them are higher than this bound. To proceed into the anisotropic regime, we derived structure equations appropriated
for a cylindrical metric with anisotropic pressures~\cite{2014arXiv1407.2280M}. From the solutions of the structure equations in cylindrical symmetry we have confirmed the same bound for $B \sim 10^{13} $~G, since beyond this value no physical solutions are possible. Our tentative conclusion is that massive WD, with masses well beyond the Chandrasekhar limit do not constitute stable solutions and should not exist.
\end{abstract}


\maketitle


\section{Introduction}

Motivated by the observation of thermonuclear supernovae that seem to require exploding white
dwarf masses above the celebrated Chandrasekhar limit~\cite{1931ApJ....74...81C}, a series of papers by
Mukhopadhyay and collaborators~\cite{2012IJMPD..2142001D,2013PhRvL.110g1102D} explored the magnetized version
of the stellar structure and argued for a substantial increase of the maximum possible mass for
large values of the magnetic field $B$, which quantizes the electronic energy levels.
A great deal of interest followed this suggestion, which has been
addressed in a number of works. The main criticisms include an inconsistency with the virial theorem~\cite{2014ApJ...794...86C} for the achievement of large maximum mass values and similar basic facts. However, the authors have responded to these criticisms, and the issue of the existence
of these compact stars is still an open and important question.

On the theoretical point of view, the construction of fully consistent equilibrium solutions in the
magnetized regime is still lacking, although hints for their existence and stability have been pointed out~\cite{2013PhRvL.110g1102D}. The spatial distribution of the magnetic field seems to be an important ingredient for this issue, while the behavior of matter under extreme conditions leads to a consideration of the equation of state in the Landau regime for the electron energy levels, which may change the effective description in terms of polytropic indexes and related quantities. Therefore, a step forward towards the solution of this problem would be the investigation of the stability of stellar models for highly magnetized matter and identify the threshold values of $B$ for the disappearance of stable solutions. We perform in this Letter such an analysis within a definite relativistic framework and show that, at least within these simplified models, the magnetic field admitted on theoretical grounds can not exceed $1.5 \times 10^{13}$~G. Moreover, we confirm that the maximum masses do not grow beyond the Chandrasekhar value when the magnetic pressure is properly introduced via the stress-energy tensor. A brief discussion of the theoretical situation of putative high-mass white dwarfs closes this work.

\section{Models of magnetized white dwarfs}

The scalar virial theorem~\cite{Shapiro1991ApJ} has been generally employed to estimate that the maximum magnetic field that a WD can sustain with a mass $M=1.4 M_\odot$ and a radius $R=0.005R_\odot$ is around $B_{\text{max}}\sim 10^{13}$~G. This values strongly suggest that a realistic model of a magnetized WD should feature quantized energy levels for the electrons, as in many attempts to construct models that describe the
microphysics of WDs as a magnetized fermion systems~\cite{Felipe2005ChJAA}.

In the approximation in which the magnetic field is constant and matter is allowed to settle in it, a
breaking of the spherical symmetry of the star is apparent. This is not very relevant for low magnetic fields, but since
we want to reach the extreme anisotropic regime, we have chosen to work in cylindrical coordinates in
which the polar and equatorial radii differ and the deviation of spherical symmetry is naturally accounted
for. An additional advantage of this procedure is that the construction of an anisotropic energy momentum
tensor for the magnetized matter is very well-defined and straightforward.

In~\cite{2014arXiv1407.2280M} we have developed a first attempt to investigate this problem using a general
cylindrical symmetric metric, whit coordinates ($t,r,\phi,z$). We have followed there the procedures of~\cite{Trendafilova2011EJPh} to solve Einstein equations for an axisymmetric model of a WD to take into
account the anisotropy induced by the magnetic field. A constant magnetic field in (say) z-direction defines
two main directions in space, parallel and perpendicular to the magnetic field. The main approximation performed
in that paper is to assume that all the functions and variables depend only on the radial coordinates
$(r)$ and not on $z$ and $\phi$, so that we can solve the dependence on the equatorial direction of the WD.
However, this simple model could be useful to obtain information of the effects of the magnetic
field in terms of the shape (oblateness) of the WD and yield upper limits for the values of the magnetic field
that this objects can sustain.

The present paper builds on the~\cite{2014arXiv1407.2280M} results, and applies the same procedure
to study the structure equation of a magnetized WD with the aim to confirm or dismiss the recently claims
of super-Chandrasekhar masses for magnetized WD~\cite{2012IJMPD..2142001D}.

\section{Magnetized white dwarfs}\label{sec1}
The thermodynamical properties of matter in a magnetic field are obtained starting from the thermodynamical potential at zero temperature
\begin{eqnarray}\label{OM}
   \Omega_e  = -\frac{eB}{4\pi ^2}\left [\sum_{l=0}^{l_{max}}\alpha_{l}\left( \mu_e\,{p}_{F}-{\varepsilon}_{e}^2\ln\frac{ {\mu_e} + {p}_{F}}{{\varepsilon}_{e}}\right)\right ],
\end{eqnarray}
where $\mu_e$ is the electron chemical potential, $l_{max}= [\frac{\mu_e^2-m_e^2}{2eB}]$, $I[z]$ denotes the integer part of $z$, $\alpha_{l}=2-\delta_{l0}$ is the spin degeneracy of the $l$-Landau level, the Fermi momentum is ${p}_F=\sqrt{{\mu_e}^2-{\varepsilon}_{e}^2}$ and the rest energy is given by
\begin{equation}
\varepsilon_{e}=\sqrt{m_e^2 +2|e B|l},
\end{equation}

The particle number density and the magnetization are
\begin{eqnarray}
N_e&=&-(\partial\Omega_e/\partial \mu_e)=\frac{m^2}{2\pi^2}\frac{B}{B^c_{e}}\sum_{l=0}^{l_{max}}\alpha_{l}{p}_F,\label{Density1}\\
\mathcal{M}_e&=&-(\partial\Omega_e/\partial B)=\frac{e m_e}{4\pi ^2}\left (\sum_{l=0}^{l_{max}}\alpha_{l}\left[{\mu_e}{p}_F-\left[ {\varepsilon}_{e}^2 + 2{\varepsilon}_{e} {C}_e \right]\ln\frac{{\mu_e}+ {p}_F}{{\varepsilon}_{e}}\right]\right) \label{Magnetizacion}
\end{eqnarray}
where $B^{c}_e=m_{e}^2/|e|=4\times10^{13}$~G is the critical
magnetic field  and ${C}_e =\frac{B}{B_e^c}l/\sqrt{2l\frac{B}{B_e^{c}}+m_e^2}$.

The energy density and the pressures parallel and perpendicular to the magnetic field can be written as
\begin{subequations}
\begin{eqnarray}
   \epsilon&=& \Omega_e + \mu_e N_e +N m_N\frac{A}{Z},\label{EnerPresure1a}\\
  \mathcal{P}_{\parallel}  &=&-\Omega_e,\label{EnerPresure1b}\\
   \mathcal{P}_\bot &=&-\Omega_e-B\mathcal{M}_e,\label{EnerPresure1c}
\end{eqnarray}
\end{subequations}
where $N m_N\frac{A}{Z}$ is the mass density term, $N$ is the number of nucleons, $m_N$ is the mass of nucleons, Z is the atomic number and A baryon number. We assume that the white dwarfs are predominantly composed by $\,^{12}C$ and $\,^{16}O$ with $A/Z = 2$.

The EoS including matter and field ($P_{\perp}^B=E^B=-P_{\parallel}^B=\frac{B^2}{8\pi}$) contributions have the following form
\begin{eqnarray}
  E &=& \varepsilon+\frac{B^2}{8\pi}, \label{EoS1}\\
  P_{\parallel} &=& \mathcal{P}_{\parallel}-\frac{B^2}{8\pi}, \label{EoS2}\\
  P_{\perp} &=& \mathcal{P}_{\perp}+\frac{B^2}{8\pi}.\label{EoS3}
\end{eqnarray}

\begin{figure}[!ht]
\centering
      \includegraphics[height=10.0cm,width=12cm]{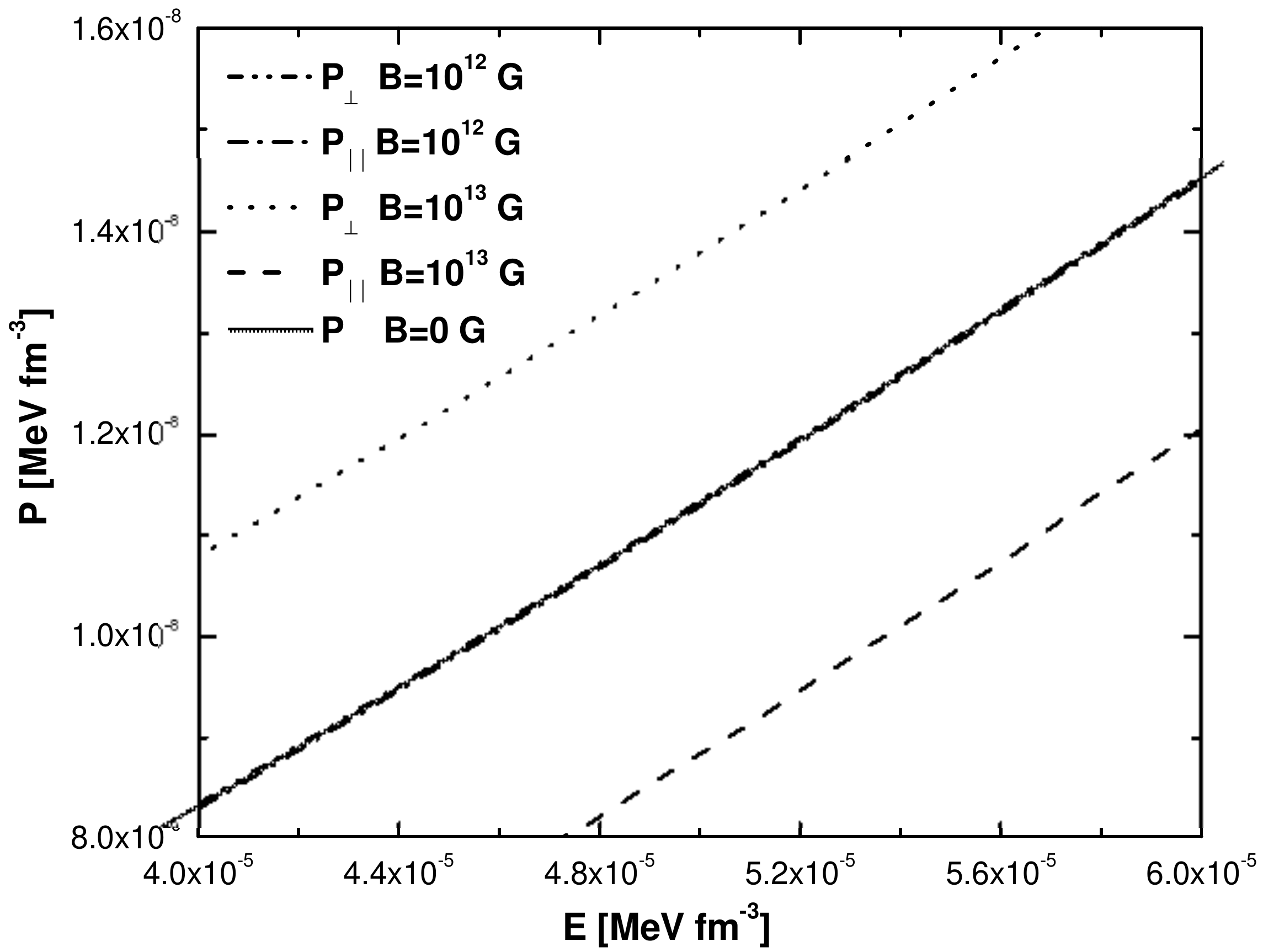}
      \caption{Equations of state for magnetized electron gas. Notice the differences of the pressures when the magnetic field increase, the cases of $B=0$ and $B=10^{12}$ almost can not be distinguished one of the other. }
      \label{fig1}
\end{figure}

In Figure~\ref{fig1} we show the EOS of the magnetized gas as derived from the above expressions. It can be noticed that for low values of the magnetic field ($B \ll B_e^c$), the pressures do not differ much among them
and from the non-magnetic case $B = 0$. However, when $B \sim B_e^c$, the difference of the pressures
becomes quite large.

\begin{figure}[!ht]
\centering
      \includegraphics[height=10.0cm,width=12cm]{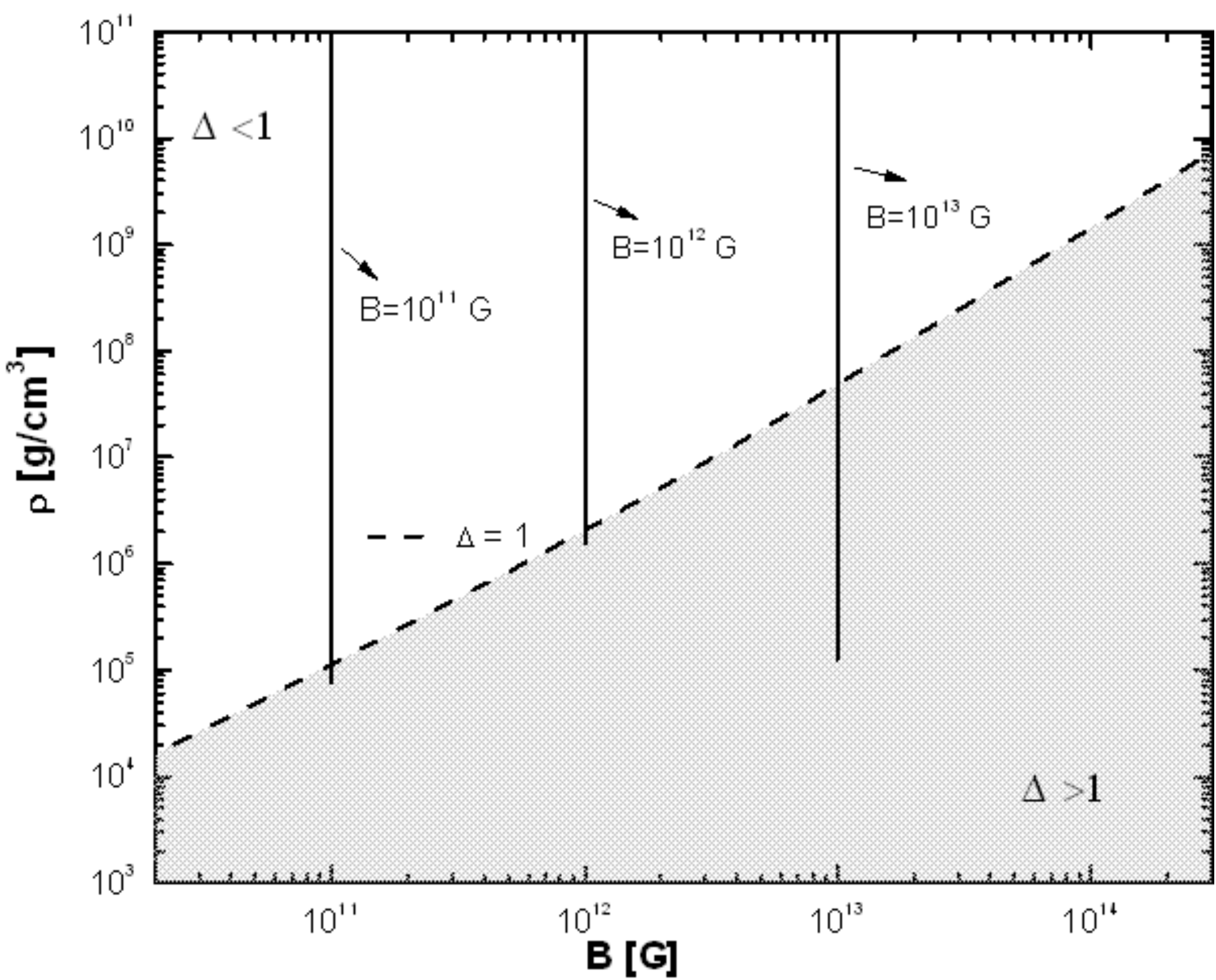}
      \caption{Splitting of the pressures whit respect to the magnetic field. The dashed line represent the solution of the equation  $\Delta(\mu_e,B)=1$ and the vertical lines are solutions of the densities for a constant magnetic field value.}
      \label{fig22}
\end{figure}

To quantify the anisotropy  we have defined the splitting coefficient as
\begin{equation}\label{split}
    \Delta=\frac{|P_{\perp}-P_{\parallel}|}{P(B\rightarrow0)}.
\end{equation}

We will use as a criterion that the border separating the isotropic and the anisotropic regions is just
$\Delta\simeq\mathcal{O}(1)$. We have solved the equation $\Delta(\mu_e,B)=1$ in order to distinguish the anisotropic region from the isotropic one. In Figure~\ref{fig22} we show the plane of matter densities as a function of the magnetic field, the region above the dashed curve falls in the isotropic regime ($\Delta<1$) and the grey region below the curve is in the anisotropic regime ($\Delta>1$). The vertical lines are the solutions of the densities for a constant magnetic field value, for $B\lesssim10^{12}$~G, all the points lay in the isotropic region while for $B\gtrsim10^{13}$~G there is a considerable number of points in the anisotropic region.

In our numerical computations we will first use magnetic field values well within the isotropic region $B=10^{12}$~G and also in the anisotropic region $B=10^{13}$~G to compare the effects on the star structure. In next section it will  be shown the impact on the stellar structure of this density-dependent field anisotropy.

\section{TOV equations for Magnetized  White Dwarfs}\label{sec2}

In order to set up the problem introduced by the magnetized matter EoS in the study of the structure
of WDs, we will analyze the usual case first, assuming spherical symmetry and solving the resulting TOV
equations~\cite{Misner1973grav}. To find the static structure of a relativistic spherical star we have to solve the well-known TOV equations.
\begin{eqnarray}
\frac{dM}{dr} &=& 4\pi G\,E, \label{TOV1} \\
\frac{dP}{dr} &=& -G\frac{(E+P)(M+4\pi Pr^3)}{r^2-2rM}, \label{TOV2}
\end{eqnarray}
with the boundary conditions $P(R)=0$, $M(0)=0$ and the EoS $E\rightarrow f(P)$.

The Mass-Radius curves obtained for magnetic fields $B=10^{11},10^{12}$~G (that is, within the regime in
which $\Delta < 1$) are in agreement with the ones obtained, for example, in~\cite{2000ApJ...530..949S}. Therefore, we confirm that quantization of the electronic levels for fields in this regime can not increase the maximum mass of a WD sequence. However, if the magnetic field pressure terms $\propto B^{2}/8 \pi$ are omitted, higher values can be achieved. We believe that this unjustified omission is a significative part of the discussion on super-
Chandrasekhar masses.

When we want to explore the anisotropic region, however, we find that the anisotropy sets in for progressively
large regions of the WD when the value of the magnetic field is increased. Around $B = 10^{13}$~G most
of the star “feels” the anisotropy of the pressures and just the inner regions remain practically isotropic.
This justifies the use of anisotropic solutions for the highest fields.

\section{Anisotropic structure equations}\label{sec3}

In order to improve the structure equations in presence of anisotropic pressures we consider in this section an axisymmetric geometry which is
more adequate to treat magnetized fermion system. We follow the same procedure as in~\cite{2014arXiv1407.2280M}. The cylindrically symmetric metric reads
\begin{equation}\label{cyl1}
  ds^2=-e^{2\Phi} dt^2+e^{2\Lambda} dr^2+r^2d\phi^2+e^{2\Psi} dz^2
\end{equation}
where $\Phi$, $\Lambda$, $\Omega$, and $\Psi$ are functions of $r$ only which as mention before is a main approximation~\cite{2014arXiv1407.2280M}.

With the energy momentum tensor for magnetized mater given by~\cite{Felipe2005ChJAA}
\begin{equation} \mathcal{T}^{\mu}_{\,\,\,\,\nu} = \left(
\begin{array}{llll}
E & 0&0&0 \\
0 & P_{\perp}&0&0\\
0&0&P_{\perp}&0\\
0&0&0&P_{\parallel}
\end{array}\right), \label{presiones}
\end{equation}
where $E$, $P_{\parallel}$ and $P_{\perp}$ are given by the EoS (\ref{EoS1}), (\ref{EoS2}) and (\ref{EoS3}) respectively.

From the Einstein field equations in natural units  and using the energy momentum conservation $(\mathcal{T}^{\mu}_{\,\,\,\,\nu;\mu})$, we obtain the following four differential equations:
\begin{subequations}\label{Diff2}
\begin{eqnarray}
P_{\perp}'&=&-\Phi'(E+P_{\perp})-\Psi'(P_{\perp}-P_{\parallel}),\\
4\pi e^{2\Lambda}(E+P_{\parallel}+2P_{\perp})&=&\Phi''+\Phi'(\Psi'+\Phi'-\Lambda')+\frac{\Phi'}{r}, \\
4\pi e^{2\Lambda}(E+P_{\parallel}-2P_{\perp})&=&-\Psi''-\Psi'(\Psi'+\Phi'-\Lambda')-\frac{\Psi'}{r}, \\
4\pi e^{2\Lambda}(P_{\parallel}-E)&=&\frac{1}{r}(\Psi'+\Phi'-\Lambda').
\end{eqnarray}
\end{subequations}
This, together with the EoS $E\rightarrow f(P_{\perp}),\,\, P_{\parallel}\rightarrow f(E)$ is a system of differential equations in the variables
\begin{equation}
 P_{\perp},\,\,\,P_{\parallel},\,\,\,E,\,\,\, \Phi,\,\,\, \Lambda,\,\,\, \Psi,
\end{equation}

\begin{figure}[!ht]
\centering
      \includegraphics[height=10.0cm,width=12cm]{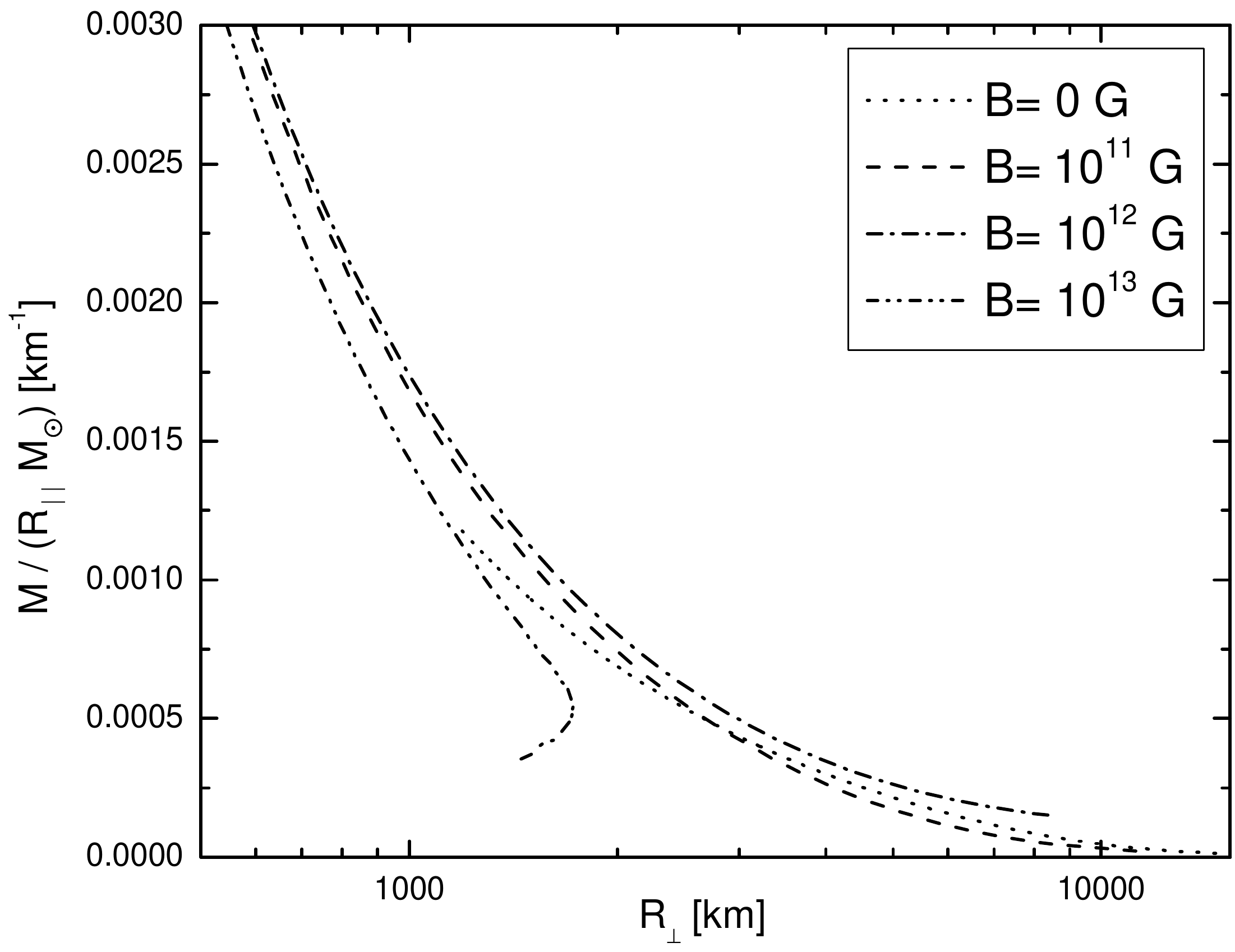}
      \caption{Mass per unit parallel length $(M/R_{\parallel})$ in solar mases, versus perpendicular radius. We have plotted curves for  the following magnetic fields values: $B=0$~G (isotropic case), $B=10^{11}$~G, $5\times10^{11}$~G, $10^{12}$~G   $5\times10^{12}$~G and $B=10^{13}$~G. This last  value of the magnetic field is the maximum value that we found stable configurations.}
      \label{fig55}
 \end{figure}

We consider $P_{\perp}(R_{\perp})=0$ which determines the radii of the star, in the equatorial (perpendicular) direction.
The solution of the system of equations (\ref{Diff2}) are shown in Figure~\ref{fig55}.

We have plotted in Figure~\ref{fig55} the mass per unit length versus perpendicular radius. (we can not
calculate a “standard” Mass-Radius relation because we did not retain a dependence of our magnitudes
on the $z$-direction for this choice of the metric). When the magnetic field increases, the perpendicular radius
and the mass per unit length of the star also increase. We have also found that there is a maximum field
($B\simeq10^{13}$~G) beyond which the metric coefficients exhibit a divergent behavior, this value of the magnetic field coincides with the value when the splitting of the pressures given by the parameter $\Delta$ is greater than 1, for $B=10^{13}$~G. This result supports our interpretation that beyond this value of the magnetic field there are no stable solutions of the system, and point towards the end of the theoretical stellar sequences constructed from our assumptions.

\section{Conclusions}\label{sec4}

We have revisited the role of anisotropic pressures in the description of the structure of a WD. Our findings
show that when the splitting coefficient $\Delta > 1$, the differences in the pressures can not be neglected and a different approach must be used to study the structure of the star. An axisymmetric geometry is more suitable than a spherical one for the solution of Einstein equation using a cylindrical symmetric metric. Our choice of the
metric in the latter conditions is probably the simplest among all possible cylindrical metrics.

Taking into account the pressure anisotropy due to a magnetic field yields a critical field $B_{c}\sim10^{13}$~G for magnetized WD, beyond which there are no stable equilibrium configurations. This bound for the value of the magnetic field is close (but slightly lower) to the one obtained based on the scalar virial theorem.

Although in our model we can not compute the total mass due to the assumption that all the variables
depend only on the perpendicular (equatorial) radius and not of the $z$-direction~\cite{2014arXiv1407.2280M}, the study is useful to confirm the existence of a maximum magnetic field for which the star may
undergo an anisotropic collapse due to a magnetic instability. This point helps to clarify the claim of super-
Chandrasekhar masses for magnetized WD~\cite{2012IJMPD..2142001D,2013PhRvL.110g1102D} and rules out that the
magnetic field could be the reason of the existence of this kind of objects. By the way, the recent paper~\cite{2014arXiv1411.5367D} makes use of extremely high magnetic fields, well above the Schwinger value, and clearly beyond the virial estimate; and also imposes a $\Gamma = 4/3$ polytrope as a model for the matter. This is at odds with previous claims ~\cite{2013PhRvL.110g1102D} by the same authors that a $\Gamma = 2$ results
from Landau quantization and hence it is clear that the latter does not stiffen the equation of state
needed to achieve the super-Chandrasekhar mass values.

\acknowledgments
The authors  thanks to H. Quevedo, R. Pican\c co, Kepler da Oliveira, J. Rueda and R. Ruffini for fruitful comments and discussions. The work of A.P.M and D.M.P. have been supported  under the grant CB0407 and the ICTP Office of External Activities through NET-35.
D. M. P acknowledgment the fellowship CLAF-ICTP  and also thanks to IGA-USP for the hospitality.
A.P.M thanks the hospitality and support given by the International Center for Relativistic Astrophysics Network,
specially to Prof Remo Ruffini where this paper has been done. JEH wishes to thank the financial support of the CNPq and FAPESP Agencies (Brazil).

\label{lastpage}

\end{document}